\documentclass[preprint,showpacs,eqsecnum,floatfix]{revtex4}
\usepackage{amsmath}
\usepackage{amssymb}
\usepackage{amsfonts}
\usepackage{graphicx}
\def\beq{\begin{equation}}
\def\eeq{\end{equation}}

\def\p{{\sf p}}

\def\x{{\sf x}}

\def\q{{\sf q}}

\def\y{{\sf y}}

\def\a{{\sf a}}

\def\X{{\sf X}}

\begin{document}
\title{Scattering and reflection positivity in relativistic 
Euclidean quantum mechanics}
\author{W. N. Polyzou}
\affiliation{
The University of Iowa, Iowa City, IA
52242}

\begin{abstract}

  In this paper I exhibit a class of reflection positive Euclidean
  invariant four-point functions that can be used formulate a
  Poincar\'e invariant quantum theory.  I demonstrate the existence of
  scattering wave operators, which can be calculated without analytic
  continuation in this representation.


\end{abstract}

\pacs{03.70.+k, 11.10.-z 11.15.Ha 64.60.ae}

\maketitle

\section{Introduction}

In this paper I discuss the existence of scattering wave operators in
a representation of relativistic quantum mechanics where the dynamics
is introduced through a collection of reflection positive Euclidean
``Green functions''.

This work is motivated by the observation that the physical Hilbert
space inner product in the Osterwalder-Schrader reconstruction theorem
of Euclidean field theory \cite{Osterwalder:1973dx} can be expressed
as a sesquilinear form involving only Euclidean Green functions and
Euclidean test functions.  Furthermore, the properties of the
Euclidean Green functions that are needed to reconstruct a
relativistic quantum theory are subset of the properties needed to
construct a local field theory.  There are explicit representations of
all ten Poincar\'e generators as self-adjoint operators in this
Euclidean representation of the Hilbert space.  The important 
observation is that this representation of the relativistic 
quantum theory does {\it not} require analytic continuation.

The general formalism was presented in a previous paper
\cite{Kopp:2011vv}.  In that paper we introduced methods to compute
sharp-momentum transition operators using matrix elements of
$e^{-\beta H}$ in normalizable states.  These matrix elements can be
computed using only quadratures in the Euclidean representation.  The
computational methods that we introduced were tested using a solvable
model and convergence to the exact transition matrix elements was
demonstrated for energies between 50 MeV and 2 GeV.

The dynamical input is a set of multi-point Euclidean Green functions
that are defined as moments of Euclidean path integrals or
solutions of Schwinger-Dyson equations.  Actual computations use
finite collections of approximate or model Green functions.  The
Euclidean Green functions that define an acceptable quantum theory
satisfy a condition called reflection positivity, which is equivalent
to the requirement that vectors in the Euclidean representation of the
Hilbert space have non-negative norm.

Reflection positivity in continuum theories is a restrictive
condition.  This is expected since it ensures the existence of an 
analytic continuation of spacetime variables to Minkowski space. 
It is difficult to check because the sum of a reflection
positive Green function and a small Euclidean invariant perturbation
is not necessarily reflection positive \cite{vic}.  In addition, products of
reflection positive operators are not necessarily reflection positive.
These observations imply that the solution of the Bethe-Salpeter
equation with reflection positive input is not automatically
reflection positive.  A natural question that arises is are there real 
fundamental problems with finding reflection positive Green functions,
or do the issues discussed above just make it difficult to demonstrate? 
Having a class of non-trivial examples would make it clear that there are
no fundamental problems in constructing reflection positive 
Green functions. 

This paper addresses two questions which were not addressed in
\cite{Kopp:2011vv}.  The first question, discussed in the previous
paragraph, is ``Are there any non-trivial reflection-positive
multipoint Green functions?''  The second question, assuming that
first question is answered in the affirmative, is ``Do these
reflection positive Green functions lead to a non-trivial scattering
theory?''  This is the Euclidean version of determining what class of
potentials lead to a non-trivial scattering in non-relativistic
quantum mechanics.

This paper does not provide a general solution to this problem, but it
exhibits a representative class of reflection positive Euclidean
invariant four-point functions that are motivated by a one-dimensional
structure theorem. This paper also shows that scattering wave
operators exist for this class of four-point functions.  This paper
also includes a short discussion of issues related to applying these
methods to QCD as well as a discussion of how to extend the framework
presented in \cite{Kopp:2011vv} to treat states of any spin.



The advantages of this approach are (1) all relevant calculations can
be performed entirely in Euclidean space, (2) there is numerical
evidence that GeV-scale scattering calculations can be performed
without analytic continuation (3) and finite Poincar\'e
transformations can be performed without leaving Euclidean space.

In the next section I introduce notation and summarize needed results
from \cite{Kopp:2011vv}.  In section three I exhibit a class of reflection
positive four-point Green functions.  In section four I discuss the
formulation of the scattering problem when the dynamics is in the
kernel of the scalar product and demonstrate the existence of
scattering wave operators based on the Green functions introduced in
section three.  In section five I briefly review computational methods that
can be applied to the examples of sections three and four. Section six
provides a brief summary of the key results of this paper and a 
discussion of remaining open problems.  The appendix discusses 
the structure of two-point Euclidean Green functions for arbitrary spin.

\section{definitions and assumptions}

This section summarizes the essential background material from
\cite{Kopp:2011vv}. 

A relativistic Euclidean quantum theory is defined by a collection of
Euclidean invariant distributions, $S_{m:n}(\x_m, \cdots,
\x_1; \y_1, \cdots , \y_n)$, where the $\x_i$ and $\y_j$ are
final and initial Euclidean space-time variables.  The collection can
be finite or infinite.  

A dense set of Hilbert space vectors are represented by test functions
\beq
\{ f_m (\x_1, \cdots ,\x_m ) \}
\label{n:a1}
\eeq
with support for positive relative Euclidean times
\beq
0 < \x_1^0 < \x_2^0 < \cdots < \x_m^0 .
\label{n:a2}
\eeq
The physical Hilbert space inner product is 
\beq
\langle f\vert g\rangle  = 
\sum_{m,n} \int f_m^* (\x_1, \cdots ,\x_m ) 
S_{m:n}(\theta \x_m, \cdots, \theta \x_1; \y_1, \cdots , \y_n)
g_n (\y_1, \cdots, \y_n)d^{4m}\x d^{4n}\y ,
\label{n:a3}
\eeq
where $\theta \x = \theta (\x^0,\mathbf{x}) = (-\x^0,\mathbf{x})$
is the Euclidean time reflection operator.  The 
Hilbert space generated by the functions (\ref{n:a1}) with 
inner product (\ref{n:a3}) is denoted by ${\cal H}$.

The collection of Green functions in (\ref{n:a3}) 
is reflection positive if 
\beq
\langle f \vert f \rangle \geq 0.
\label{n:a4}
\eeq
This inner product has zero norm-vectors, so the actual 
Hilbert space vectors are equivalence classes of functions of the 
form (\ref{n:a1}-\ref{n:a2}), under the equivalence relation that the norm 
of the difference between equivalent functions is zero.

In quantum field theory microscopic locality requires that the
collection of Green functions is infinite and there is no distinction
between the initial and final variables; in relativistic quantum
mechanical models both of these conditions can be relaxed.  Relaxing
these conditions leads to violations of microscopic locality, however
all of the other observable requirements (axioms) of a relativistic
quantum theory remain satisfied \cite{Osterwalder:1973dx}. Computable
models involve finite numbers of degrees of freedom, so the full set
of requirements of the field theory will not be realized, but it is
still desirable to retain the relativistic invariance of the quantum
theory.

In addition, the collection of Green functions should be Hermitian, 
\beq
S_{k:n}(\x_k \cdots \x_1; \y_1 \cdots  \y_n) =
S_{n:k}^*(\y_n \cdots \y_1; \x_1 \cdots  \x_k) ,
\label{n:a5}
\eeq
so $\langle f\vert g \rangle = \langle g \vert f
\rangle^*$ and satisfy cluster properties,
\beq
S_{k:n} (\x_k , \cdots , \x_1; \y_1 , \cdots ,  \y_n) \to
\prod S_{k_i:n_i} (\x_{ki}, \cdots , \x_{1i}; \y_{1i}, \cdots ,  \y_{ni}),
\label{n:a6}
\eeq
as distributions, which means that they become products of 
fewer-point Green functions when different clusters of initial and 
final variables are asymptotically separated.

The Poincar\'e generators on ${\cal H}$ have simple expressions.
The Hamiltonian and square of the invariant mass operator are
\[
\langle \x \vert H \vert \mathbf{f} \rangle :=
-{d \over da} \langle \x-(a,0,0,0)  \vert \mathbf{f} \rangle_{\vert_{a=0}} =
\]
\beq
\{0 ,{\partial \over \partial
\x^0_{11}} f_1 (\x_{11}), 
\left ( {\partial \over \partial
\x^0_{21}} + {\partial \over \partial
\x^0_{22}} \right )  f_2 (\x_{21},\x_{22}), \cdots  \}
\label{n:a7}
\eeq
and
\beq
M^2 = H^2 - P^2 
\label{n:a8}
\eeq
\beq
\langle \x \vert M^2 \vert \mathbf{f} \rangle :=
\{0 , -\nabla_{11}^2 f_1 (\x_{11}), 
-(\nabla_{21} + \nabla_{22})^2
f_2 (\x_{21},\x_{22}), \cdots  \}
\label{n:a9}
\eeq
where $\nabla^2$ is the 4-dimensional Euclidean Laplacian.
The generators of spatial translations and rotations follow
in the usual way from the translational and rotational 
invariance of the Green functions.  Matrix elements of the 
form $\langle f \vert e^{-\beta H} \vert g \rangle$ are important
in applications and can be computed
by quadrature by replacing 
\beq
g_n (\y_1, \cdots, \y_n) \to 
g_n (\y_1-(\beta ,\mathbf{0}) , \cdots, \y_n-(\beta ,\mathbf{0}))
\label{n:a10}
\eeq 
in (\ref{n:a3}).

\section{Reflection positivity}

One difficulty with constructing reflection positive multipoint
functions is that there are non-trivial functions associated with zero
norm vectors.  

If a free Euclidean Green function is perturbed by adding a small
connected perturbation that is only required to be Euclidean
invariant, then a function representing a zero norm vector with
respect to the product of the free Euclidean Green functions might
have a non-zero contribution due to the perturbation.  This
contribution can always be made negative using the freedom to adjust
the sign of the perturbation.  The means that reflection positivity is
not stable with respect to small Euclidean invariant perturbations.
The practical consequence of this observation is that the solution of
the Euclidean Bethe-Salpeter equation,
\beq
S_{2:2} = S_{1:1} S_{1:1} + S_{1:1} S_{1:1} K S_{2:2},
\label{f.2}
\eeq
with a model Euclidean invariant kernel, $K$, is not automatically
reflection positive, even if the kernel is small.  I am not aware of
any general results about what kind of restrictions are needed on
Euclidean invariant Bethe-Salpeter kernels for $S_{2:2}$ to be
reflection positive.  In addition, reflection positivity
ensures the existence of an analytic continuation to 
Minkowski space in space-time variables, which suggests that 
it is a restrictive condition.    


The absence of any general methods for constructing reflection
positive Green functions is a problem if one wants to make
phenomenological models of Euclidean Green functions to use in this
framework.  Fortunately, a general structure theorem exits for the
two-point function in the one-dimensional case.  In what follows this
one-dimensional result will be used to motivate the construction of a
class of reflection positive four-point functions.


Since cluster properties imply that four-point functions can be 
expressed as a sum of products of reflection-positive two-point 
functions and a connected four-point function,
\beq
S_{2:2} = \sum S_{1:1} S_{1:1} + S^c_{2:2}, 
\label{f.4}
\eeq  
in order to show $S_{2:2}$ is reflection positive it is sufficient to
show that the connected four-point function, $S^c_{2:2}$, is
reflection positive.

In one dimension there is a result due to Widder
\cite{Widder:1941}\cite{Widder:1931}\cite{Widder:1934} from classical
analysis that gives the general structure of reflection positive
two-point functions.  Widder's theorem points out that any kernel
$k(s)$ satisfying the reflection positivity condition

\beq
\int f(\theta s) k(s-t) f(t) ds dt =
\int f(s) k(-s-t) f(t) ds dt
\geq 0
\label{f.5}
\eeq
can be expressed in the exponential form 
\beq
k(-\tau'-\tau) = \int e^{- \lambda (\tau'+\tau) } \rho (\lambda ) d\lambda
\label{f.6}
\eeq
for some positive density $\rho(\lambda)$.  Since in this example, 
$\tau',\tau>0$, we can write the kernel as
\beq
k(-\tau'-\tau) =  \int_0^{\infty} {\lambda \over \pi} 
\rho (\lambda ) d\lambda \int_{-\infty}^\infty ds  
{e^{-is ( \tau'+\tau) }\over s^2 + \lambda^2 }.  
\label{f.7}
\eeq
This has the form of a one-dimensional version of the
K\"all\'en-Lehmann representation of a two-point Euclidean
Green function.

The Widder result suggests that connected four-point 
Euclidean Green functions with the structure 
\[
S^c_{2:2} (\x_2,\x_2;\y_1,\y_2) = 
\]
\[
\int 
d^4\p_1 d^4\p_2 d^4\p_3 dm_a dm_c dm_b
e^{-i \p_1 \cdot (\x_2 -\x_1)}
e^{-i \p_2 \cdot (\x_1 -\y_1)}
e^{-i \p_3 \cdot (\y_1 -\y_2)}
\times 
\]
\beq
{ v(m_a, \p_1,m_c, \p_2, m_b, \p_3) \over 
(\p_1^2 + m_a^2 )(\p_2^2 + m_c^2) (\p_3^2 + m_b^2) 
}
\label{f.8}
\eeq
would be reflection positive for suitable Euclidean invariant kernels,
$v(m_a, \p_1,m_c, \p_2, m_b, \p_3)$.  The relevant observation is that 
this reduces to the one-dimensional case if there is enough symmetry 
between the initial and final variables.  
This structure does not provide
a general representation for a Euclidean invariant reflection positive
four point functions, as one gets in Widder's theorem.  On the other
hand, Widder's theorem suggests that reflection positivity and
Euclidean covariance strongly constrain the class of reflection
positive four-point functions.  

The contribution to the Hilbert space norm 
from connected Green functions of the form (\ref{f.8}) is
\[
\Vert \vert fg \rangle \Vert^2_c = 
\]
\[
\int 
d^4 \x_1 d^4\x_2 d^4\y_1 d^4\y_2 d^4 \p_1 d^4\p_2 d^4\p_3 dm_a dm_c dm_b 
f_e^*(-\x_2^{0}, \mathbf{x}_2)
g_e^*(-\x_1^{0}, \mathbf{x}_1)
e^{-i \p_1 \cdot (\x_2 -\x_1)}
\times 
\]
\beq
e^{-i \p_2 \cdot (\x_1 -\y_1)}
e^{-i \p_3 \cdot (\y_1 -\y_2)}
{ v(m_a, \p_1,m_c, \p_2, m_b, \p_3) \over 
(\p_1^2 + m_a^2 )(\p_2^2 + m_2^2) (\p_3^2 + m_b^2) 
}
f_e(\y_2^{0},\mathbf{y}_2)
g_e(\y_1^{0},\mathbf{y}_1)
\label{f.9}
\eeq
where the functions satisfy the support condition, 
$g_e (\y_1^{0},\mathbf{y}_1) $ and
$f_e (\y_2^{0}, \mathbf{y}_2) $  
can be non-zero only for $ 0 < \y_1^{0} < \y_2^{0}$. 

The most straightforward assumption is to choose $v(m_a, \p_1,m_2, \p_2,
m_b, \p_3)$ to be symmetric with respect to $\mathbf{p}_1 ,m_a \leftrightarrow 
\mathbf{p}_3,m_b$ and 
analytic in the upper-half $p_i^0$ planes.  In this case
the $\p^0_i$ integrals can be performed as in the one-dimensional case,
where the convergence in the upper-half plane is ensured by the Euclidean
time-support constraints provided the kernel is polynomially bounded,
and the Minkowski boundary value is a tempered distribution.
The integral over the vector variables leads to Fourier transforms of
the vector variables and results in the expression
\[
(2 \pi)^9 \int 
d\x^0 d\y^0 d\x^{0\prime} d\y^{0\prime}  d\mathbf{\p}_1 
d\mathbf{\p}_2 d\mathbf{\p}_3 dm_a dm_c dm_b 
\tilde{f}_e^*(\x_2^{0\prime },-\mathbf{p}_1 )
\tilde{g}_e^*(\x_1^{0\prime}, \mathbf{p}_1-\mathbf{p}_2) \times
\]
\[
e^{- \omega_{m_a} (\mathbf{p}_1) (\x_2^{0} - \x_1^{0}) }
e^{- \omega_{m_c} (\mathbf{p}_2) (\x_1^{0} + \y_1^{0} ) }
e^{- \omega_{m_b} (\mathbf{p}_3) (\y_2^{0} - \y_1^{0}) }
\times 
\]
\beq
{ v (m_1,(i \omega_{m_a}(\mathbf{p}_1), \mathbf{p}_1),m_c, 
(i \omega_{m_c}(\mathbf{p}_2), \mathbf{p}_2),
(i \omega_{m_b}(\mathbf{p}_3), \mathbf{p}_3),m_b)
\over 
2 \omega_{m_a} (\mathbf{p}_1)
2 \omega_{m_c} (\mathbf{p}_2)
2 \omega_{m_b} (\mathbf{p}_3) 
}
\tilde{f}_e(\y_2^0, \mathbf{p}_3-\mathbf{p}_2)
\tilde{g}_e(\y_1^0, -\mathbf{p}_3).
\label{f.10}
\eeq
The  kernel, which was initially a Euclidean invariant function of Euclidean
scalar products becomes a Lorentz invariant function of Lorentz invariant inner 
products when the residues are evaluated at the poles
$\p^0\to i \omega_{m_i}(\mathbf{p}_i)$.
This connected Euclidean Green function
will be reflection positive if
\beq
v (m_a,(i \omega_{m_a}(\mathbf{p}_1), \mathbf{p}_1),m_c, 
(i \omega_{m_c}(\mathbf{p}_2), \mathbf{p}_2),
(i \omega_{m_b}(\mathbf{p}_3), \mathbf{p}_3),m_b)
\label{f.12}
\eeq
is a positive symmetric matrix in $\mathbf{p}_1, m_a$ and
$\mathbf{p}_3, m_b$ for all values of $\mathbf{p}_2$ and $m_c$.
This is not a difficult condition to realize.

This construction demonstrates the existence of a large class
of reflection positive connected four-point Euclidean Green functions.
Exchange symmetry puts additional constraints on the Euclidean Green function
functions.  This construction provides a class of non-trivial 
reflection positive Green functions that can be used to investigate 
properties of non-trivial models.

The second question that can now be addressed is whether the relativistic
quantum models defined by these Green functions have non-trivial
scattering operators.
 
\section{Scattering theory}

In theories like quantum field theory, where the dynamics is encoded
in the kernel of the quantum mechanical scalar product, there is no
natural asymptotic dynamics on the physical Hilbert space to formulate
scattering asymptotic conditions.

Instead, in order to formulate the scattering theory, 
cluster properties of the Green functions are utilized to formulate
the asymptotic conditions.
Cluster properties imply
that in each asymptotic region the Green functions break up into
products of subsystem Green functions.  Since the Green functions
define the structure of the Hilbert space, in each asymptotic region
the Hilbert space looks like the tensor product of Hilbert spaces
associated the dynamics of asymptotically separated subsystems.  
Scattering asymptotic
conditions can be formulated by finding eigenfunctions associated with
the point spectrum of each of the subsystem mass operators using
(\ref{n:a9}).  These can be further decomposed into simultaneous
eigenstates of mass, linear momentum, spin and spin projection using
the space translations and rotations (see (4.4) equations 
and (4.6) of \cite{Kopp:2011vv}).  These vectors are a basis for an irreducible
representation space of the Poincar\'e group.  Products of these
subsystem bound-state eigenfunctions define a mapping from a Hilbert
space of scattering asymptotes, which is the tensor product of
Poincar\'e irreducible representation spaces, to ${\cal H}$.

%
%
%
In order to illustrate the main concepts I consider the special 
case of two-particle scattering.  Also for simplicity I assume
that $1$ and $2$ correspond to different types of scalar particles.
In this case the dynamics is given
by a 4-point Euclidean Green function
\beq
S_{2:2} (\x_2,\x_1:\y_1,\y_2).
\label{i.1}
\eeq
The cluster condition means that the four-point function 
has the form  
\[
S_{2:2} (\x_2,\x_1:\y_1,\y_2) =
\]
\beq
S_{1:1} (\x_1:\y_1)S_{1:1} (\x_2:\y_2) +
S^c_{2:2} (\x_2,\x_1:\y_1,\y_2)
\label{i.2}
\eeq
where $S^c_{2:2} (\x_2,\x_1:\y_1,\y_2)$ is a connected four-point function
and $S_{1:1} (\x_i:\y_i)$ are two-point functions.  

I also assume that $S_{1:1} (\x_i:\y_i)$ is the two-point Euclidean Green
function for a free scalar particle of mass $m_i$.  In this case every vector 
in the Hilbert space associated with $S_{1:1} (\x_i:\y_i)$ 
is a mass eigenstate with eigenvalue $m_i$ so there is no need to 
solve the mass eigenvalue problem discussed above.
%
%
%
These asymptotic one-particle eigenstates are represented by functions of
the Euclidean variables, 
\beq
\langle \x_i \vert m_i \rangle =  \psi_i(\pmb{\x_i}) h_i(\x_i^0) 
=
h_i(\x_i^0) \int {d \mathbf{p} \over (2 \pi)^{3/2}}
e^{i \pmb{\p} \cdot \pmb{\x}_i} \tilde{\psi}_i(\mathbf{p}) 
\eeq
where  $h_i(\x_i^0)$ is a smooth function that has compact support for 
positive Euclidean time. The freedom to choose the functions $h_i(\x_i^0)$ 
is related to the fact the vectors are represented by 
equivalence classes of functions.  Specifically, in the asymptotic
region, the different $h_i$'s are all associated with the same one-body 
Poincar\'e irreducible basis states. 
 
I define a mapping, $\Phi$ from the tensor 
product of the space of square integrable functions
of the $\pmb{\p}_i$ to the 
Hilbert space ${\cal H}$ by
\beq
\langle \x_1 ,\x_2 \vert \Phi \vert \psi_1 \otimes \psi_2 \rangle
=   
h_1(\x_1^0) h_2(\x_2^0) \int {d\pmb{\p}_1 \over (2 \pi)^{3/2}}
{d\pmb{\p}_2 \over (2 \pi)^{3/2}}
e^{i \pmb{\p}_1 \cdot \pmb{\x}_1} \tilde{\psi}_1(\pmb{\p}_1) 
e^{i \pmb{\p}_2 \cdot \pmb{\x}_2} \tilde{\psi}_2(\pmb{\p}_2) 
\eeq
where the $h_i$ can be chosen so their supports do not intersect.  Since 
the particles are spin zero particles, this defines a mapping from
the tensor product of two irreducible representation spaces
of the Poincar\'e group (mass $m_1$, spin 0 and mass $m_2$, spin 0),
${\cal H}_0$,
to the Hilbert space, ${\cal H}$.  The functions $h_1(\x_1^0) h_2(\x_2^0)$
are considered as part of the mapping.  

Scattering wave operators are mappings from the asymptotic 
Hilbert space ${\cal H}_{0}:={\cal H}_{m_1} \otimes {\cal H}_{m_2}$ 
to ${\cal H}$ 
defined by the strong limits 
\beq
\Omega_{\pm} := \lim_{t \to \pm \infty} e^{i Ht} \Phi e^{-i H_{0} t} 
\label{i.18}
\eeq
where $H_{0}= \sqrt{m_1^2 + \mathbf{p}_1^2} +
\sqrt{m_2^2 + \mathbf{p}_2^2}$ is the Hamiltonian associated with 
the asymptotically free particles.
The existence of these limits depends on properties of the
connected four-point Euclidean Green function.  This will be shown below.

Conventional methods can be used to derive
sufficient conditions for this limit to exist.  The first step is to 
write the limit as the integral of a derivative
\[
\lim_{t \to  \pm \infty} e^{i Ht} \Phi e^{-i H_{0} t} =
\Phi + 
\int_0^{\pm \infty}{d \over dt}\left (  
e^{i Ht} \Phi e^{-i H_{0} t} dt \right )   =
\]
\beq 
\Phi + i \int_0^{\pm \infty}  
e^{i Ht} (H \Phi - \Phi H_{0}) e^{-i H_{0} t} dt  . 
\label{i.19}
\eeq
A sufficient condition for the existence of the wave operator is the
strong convergence of the integral (\ref{i.19}).  A sufficient 
condition for the convergence of this integral is the Cook condition
\cite{Cook:1957}, which
exploits the unitarity of the time evolution operator 
\beq
\int_a^\infty \Vert (H \Phi - \Phi H_{0}) e^{\mp i H_{0} t} 
\vert \psi \rangle \Vert dt < \infty
\label{i.20}
\eeq
and provides a bound on the norm of the integral in (\ref{i.19}). 

The quantities appearing in (\ref{i.20}) depend on the Euclidean Green function 
functions.  In this case the integrand has the form
\[
\Vert (H \Phi - H_{0} \Phi ) e^{\mp i H_{0} t} 
\vert \psi \rangle \Vert = 
\]
\beq
( \psi_{}, e^{\pm i H_{0} t} 
(\Phi^{\dagger}H - \Phi^{\dagger}H_{0})
(\theta S_{1:1}S_{1:1}  + \theta S^c_{2:2}) 
(H \Phi- \Phi H_{0})
\psi_{} )^{1/2} 
\label{i.21}
\eeq
where $(,)$ represents the Euclidean inner product and  
I have used (\ref{i.2}) to express the four-point 
Green function as the sum of a product of two-point 
functions with a connected four-point function
that vanishes asymptotically.    
 
The second goal of this paper will be realized by
showing that 
condition (\ref{i.20}) holds for the four-point Green functions 
(\ref{f.8}), which is sufficient for the existence of the strong
limits (\ref{i.18}). To do this it is necessary to show 
\beq
\int_0^\infty  \left  (\psi,e^{\mp i H_{0} t}
( \Phi^{\dagger}H - H_0\Phi^{\dagger} 
) e^{\pm iH_0t} \theta S_{2:2} 
(H \Phi - \Phi H_0) e^{\mp iH_0t} \psi \right )^{1/2} 
dt  < 
\infty .
\label{i.22}
\eeq
In (\ref{i.22}) $H \Phi - \Phi H_0$ replaces the non-relativistic
potential.  $U_0(t)$ is the time-evolution operator for the 
asymptotically free eigenstates. 

I will argue that the integrand in (\ref{i.22}) 
falls off like $t^{-3/2}$, which is
sufficient for the Cook condition (\ref{i.20}) to be satisfied.

To see this first note that cluster properties imply that the 
Euclidean Green function is the sum of a 
product of two-point Euclidean Green functions and a connected term.  
I consider a
connected term that has the structure discussed in (\ref{f.8})
\[
S(\x_1,\x_2;\x_3,\x_4)= 
\]
\beq
\int d^4{\q}_1 d^4{\q}_2 d^4 {\q}_3 dm_a dm_c dm_b 
{e^{i \q_1 \cdot (\x_1 -\x_2) + i \q_2 \cdot (\x_2-\x_3) 
+ i \q_3 \cdot (\x_3-\x_4 ) } \over
(\q_1^2 + m_a^2 ) (\q_2^2 + m_c^2)(\q_3^2 + m_b^2)}
v(\q_1,m_a,\q_2,m_c,\q_3,m_b) 
\label{i.24}
\eeq
where $m_c > m_a,m_b$ and the spectrum of $m_c$ should be continuous
for scattering.  I have already established that 
this connected Euclidean Green function is reflection positive for suitable 
$v(\q_1,m_a,\q_2,m_c,\q_3,m_b)$. 


In this case the 
asymptotic states $\Phi \vert e^{\mp i H_{0} t}
\vert \psi
\rangle $ have the 
form 
\beq
\langle \x_1, \x_2 \vert \Phi \vert e^{\mp i H_{0} t} \psi \rangle = 
\prod_{i=1}^2 h_i(\x^0_i) \int {d\mathbf{p}_i \over 
(2 \pi)^{3/2}} \psi_i(\mathbf{p}_i) e^{i \mathbf{p}_i \cdot \mathbf{x}_i \mp
i \omega_{m_i} (\mathbf{p}_i)t} 
\label{i.25}
\eeq
where I choose the functions $h_i(\x^0_i)$ to be sharply peaked with 
compact support and integrate to 1.

With this choice of wave packets
\[
\langle \x_1,\x_2 \vert 
(H \Phi - \Phi H_0)e^{\mp iH_0t}\vert  \psi \rangle =
\]
\[
(2 \pi)^3 \left ({\partial \over \partial \x^0_1} +
{\partial \over \partial \x^0_2} - \omega_{m_1} (\mathbf{p}_1) -
\omega_{m_2}(\mathbf{p}_2) \right )   
h_1(\x^0_1)h_2(\x^0_2) \times 
\]
\beq
\int {d\mathbf{p}_1 d\mathbf{p}_2 \over 
(2 \pi)^3} \psi_1(\mathbf{p}_1)\psi_2(\mathbf{p}_2) 
e^{i \mathbf{p}_1 \cdot \mathbf{x}_1 \mp i \omega_{m_1} (\mathbf{p}_1)t 
+i \mathbf{p}_2 \cdot \mathbf{x}_2 \mp i \omega_{m_2} (\mathbf{p}_2)t} .
\label{i.26}
\eeq
The integrand in the Cook condition uses (\ref{i.24}) and (\ref{i.26})  
in 
\beq
(\psi , e^{\mp iH_0t} (\Phi^{\dagger}H - H_0\Phi^{\dagger}) \Theta 
(S_{1:1} S_{1:1}+
S_{2:2}^c)
( H\Phi - \Phi H_0)e^{\pm iH_0 t} \psi ).
\label{i.27}
\eeq
In this expression the partial derivatives in (\ref{i.26}) with
respect to the Euclidean times can be integrated by parts. 
When the Euclidean time derivatives act 
on the product of the two-point functions
$S_{1:1} S_{1:1}$ they  generate energy factors (
see equations (6.4) and (6.5) of \cite{Kopp:2011vv})
that exactly cancel the asymptotic energy factors in (\ref{i.26}), 
making the $S_{1:1}S_{1:1}$ terms in (\ref{i.27}) vanish.

This means that only the connected part of the four-point
Euclidean Green function contributes to the integral (\ref{i.27}).
Thus, the connected part of the four-point function plays an analogous
role to the interaction in the non-relativistic case.

What remains has the form 
\[
(2 \pi )^{6}
\int d^4{\q}_1 d^4{\q}_2 d^4 {\q}_3 dm_a dm_c dm_b
d\x^0_1 d\x^0_2 d\x^0_3 d\x^0_4 
\left (\omega_{m_c} (\mathbf{q}_2)  - \omega_{m_1} (\mathbf{q}_1) -
\omega_{m_2}(\mathbf{q}_2-\mathbf{q}_1) \right ) \times
\]
\[   
h_1(\x^0_1)h_2(\x^0_2)
\psi^*_1(\mathbf{q}_1)
\psi_2^*(\mathbf{q}_2-\mathbf{q}_1)
e^{ \pm i \omega_{m_1} (\mathbf{q}_1)t  \pm i 
\omega_{m_2} (\mathbf{q}_2-\mathbf{q}_1)t} 
\times
\]
\[
{e^{- \omega_{m_a} (\mathbf{q}_1) (\x^0_1 -\x^0_2)  -\omega_{m_c}(\mathbf{q}_2) 
\cdot (\x^0_2+\x^0_3) 
- \omega_{m_b} (\mathbf{q}_3 ) (\x^0_3-\x^0_4 ) } \over
(\q_1^2 + m_a^2 ) (\q_2^2 + m_c^2)(\q_3^2 + m_b^2)} \times
\]
\[
e^{ \pm i \omega_{m_1} (\mathbf{q}_3)t  \mp i 
\omega_{m_2} (\mathbf{q}_2-\mathbf{q}_3)t} 
v(\q_1,m_a,\q_2,m_c,\q_3,m_b) \times
\]
\beq
\left (\omega_{m_c} (\mathbf{q}_2) - \omega_{m_1} (\mathbf{q}_3) -
\omega_{m_2}(\mathbf{q}_2-\mathbf{q}_3) \right )   
h_1(\x^0_3)h_2(\x^0_4)  
\psi_1(\mathbf{q}_3)\psi_2(\mathbf{q}_2-\mathbf{q}_3) .
\label{i.27b}
\eeq
The large-time behavior of this integral is relevant for the Cook condition.  
To estimate the large time behavior write 
\[
- \omega_{m_a} (\mathbf{q}_1) (\x^0_1 -\x^0_2)
\pm i \omega_{m_1} (\mathbf{q}_1)t =
\]
\[
- (\omega_{m_a} (\mathbf{q}_1)
- \omega_{m_1} (\mathbf{q}_1)) (\x^0_1 -\x^0_2)
\]
\beq
- \omega_{m_1} (\mathbf{q}_1) (\mp i t +  (\x^0_1 -\x^0_2))
\label{i.27c}
\eeq
and 
\[
- \omega_{m_b} (\mathbf{q}_3) (\x^0_3 -\x^0_2)
\pm i \omega_{m_1} (\mathbf{q}_3)t =
\]
\[
- (\omega_{m_b} (\mathbf{q}_3)
- \omega_{m_1} (\mathbf{q}_3)) (\x^0_3 -\x^0_2)
\]
\beq
- \omega_{m_3} (\mathbf{q}_3)( \mp i t +  (\x^0_1 -\x^0_2)) .
\label{i.27d}
\eeq
We assume that the $m_a, m_b \geq m_1$, which is the easiest 
case to consider.  These assumptions ensure that the integrals
are all convergent.

To put (\ref{i.27}) in a manageable form I make some simplifying assumptions.
First I assume that
the $h_i(\x^0)$ are sharply peaked to factor the integrand out of the
integral.  The resulting approximation leads to 
a constant multiplied by the integrand 
evaluated at  points at $\x^0_i$ in the support of $h_i(\x^0)$.  
Similarly, I change
to total and relative momentum variables and use the translational
invariance to eliminate the center of momentum degrees of freedom.
I assume that the total 3-momentum support of the wave
functions is near zero.  One total momentum integral is eliminated by
the momentum conserving delta function.  The other total momentum 
integral is approximated by setting the total momentum 
to zero and multiplying by the volume of the support of the 
total momentum.
What remains, up to a multiplicative constant, has the form
\[
\left (\psi ,U_0(\pm t)( H\Phi - \Phi H_0) U^{\dagger}_0(t) \psi ,\theta S^c_{2:2} 
(H \Phi - \Phi H_0)U_0(\mp t) \psi \right ) \to
\]
\[
\int \left ( 
\omega_{m_c} ( \mathbf{0}) -
\omega_{m_1}(\mathbf{k}') -
\omega_{m_2}(\mathbf{k}') \right )
\times 
\]
\[   
\psi_3^*(- \mathbf{k}' )\psi^*_4 
(\mathbf{k}' ) 
e^{\pm i \omega_{m_2} (\mathbf{k}')t
\pm i \omega_{m_1} (\mathbf{k}')t}
\]
\[
e^{-\omega_{m_a} (\mathbf{k}') (\x_4^0 - \x_3^0)
-\omega_{m_c} (\mathbf{0} )(\x_3^0 + \x_1^0) 
-\omega_{m_b} (\mathbf{k}) ( \x_1^0 - \x_2^0) }
\]
\[
{d\mathbf{k} d\mathbf{k}'
\over \omega_{m_a} (\mathbf{k}' )
\omega_{m_c} (\mathbf{0})
\omega_{m_b} (\mathbf{k})} 
\]
\[
\left (
\omega_{m_c} (\mathbf{0}) - \omega_{m_1} (\mathbf{k}) -
\omega_{m_2}(\mathbf{k}) \right )    
\times
\]
\beq
\psi_1(\mathbf{k})\psi_2(
-\mathbf{k}) 
e^{ \mp i \omega_{m_1} (\mathbf{k})t  
\mp i \omega_{m_2} (\mathbf{k})t}. 
\label{i.28}
\eeq 
where $\mathbf{k}$ is the momentum of one of the particles in the zero 
momentum frame.

The time dependence in this expression comes from the $\mathbf{k}$ and 
$\mathbf{k}'$ integrals.  If I use (\ref{i.27c}-\ref{i.27d}) in this
expression, assuming that $m_a\geq m_1$ and $m_b \geq m_2$ the integral 
$(\ref{i.28})$ has the general form
\beq
\int 
{d \mathbf{k} \over \omega_{m_1} (\mathbf{k})}
g(\mathbf{k}) 
e^{-\omega_{m_1} (\mathbf{k}^2) ( \x_1^0 - \x_2^0 \pm 2it ) }
\label{i.29}
\eeq
where $g(\mathbf{k})$ is a well-behaved function of $\mathbf{k}$.

It follows from the lemma on page 157 of \cite{Ruelle:1962} that, for the 
case that the wave functions $\psi_i(\mathbf{k})$ are smooth with compact 
support, integrals of this form fall off like $t^{-3/2}$ for large time.

This shows that the Cook condition (\ref{i.20}) is satisfied for
the reflection positive Euclidean Green functions of the form
(\ref{f.8}).  The asymptotic large-time behavior is identical to
the behavior found in non-relativistic scattering theory.

The relativistic invariance of the $S$ matrix can be established 
using similar methods.  The required condition in terms of the
wave operators are the intertwining conditions
\beq
U(\Lambda ,a)\Omega_{\pm} = \Omega_{\pm}U_{0}(\Lambda ,a)
\label{i.33}
\eeq
for both asymptotic conditions.  For the space translations and rotations
this condition is a consequence of the translational and rotational 
invariance of the injection operators, $\Phi$.  
For time translations this 
follows from the existence of the wave operators.     For the 
boosts a sufficient condition is
\beq
\lim_{t \to \pm \infty} 
\Vert (\mathbf{K}\Phi - \Phi \mathbf{K}_{0}) 
e^{\mp i H_{0} t} 
\vert \psi \rangle \Vert =0 
\label{i.34}
\eeq
where expressions for the boost generators are given in
\cite{Kopp:2011vv}.
As in the scattering case, the non-zero contributions to this expression 
before taking the limit come from $S^c_{2:2}$  (see \ref{i.21}).
In the two-body example above, this depends on properties 
of the connected four-point function.  When (\ref{i.34})  holds finite 
Lorentz transformations on the scattering eigenstates 
can be realized by transforming the asymptotic states using (\ref{i.33}).

\section{Computational issues}

In this paper we have demonstrated the existence of a class of
reflection positive Euclidean Green functions and shown that this
class of Euclidean Green functions leads to non-trivial scattering
operators.  The scattering operators were constructed using
conventional time-dependent methods, where cluster
properties of the Green functions were used to formulate the scattering
asymptotic conditions.

The result is that having established the existence of wave
operators and knowing how to compute matrix elements of
$e^{-\beta H}$ in a basis of normalizable states (see \ref{n:a10}),  there 
is enough information to compute transition matrix elements.

The strategy adopted in \cite{Kopp:2011vv} to perform this
computation utilized three controlled approximations.  The first is to
use narrow wave packets to extract sharp momentum transition matrix
elements from $S$-matrix elements
\beq
\langle \mathbf{p}_1', \cdots ,\mathbf{p}'_{n_\alpha} \vert t(E_\gamma +i0) 
\vert 
\mathbf{p}_1, \cdots ,\mathbf{p}_{n_\gamma}  \rangle  \approx
{i \over 2 \pi} { \langle \psi_{\alpha f} \vert (S-I) 
\vert \psi_{\gamma i} \rangle \over \langle \psi_{\alpha f} 
\vert \delta (E_\alpha - E_\gamma ) \vert \psi_{\gamma i} \rangle}.
\label{j.2}
\eeq
The convergence of these approximations is determined by the 
smoothness of the transition matrix elements.

The $S$ matrix elements needed as input 
were expressed, using the invariance 
principle \cite{kato:1966}\cite{Chandler:1976},  as matrix 
elements of $e^{i 2n e^{-\beta H}}$ in normalizable 
eigenstates 
\beq
\langle \psi_{\alpha f} \vert S \vert \psi_{\gamma i} \rangle =
\lim_{n \to \infty} \langle \psi_{\alpha f} \vert 
e^{-i n e^{-\beta H_\alpha}} \Phi_{\alpha}^{\dagger}e^{2ine^{-\beta H}} 
\Phi_{\gamma} 
e^{-in e^{-\beta H_\gamma}} \vert \psi_{\gamma_i} \rangle . 
\label{j.3}
\eeq
The convergence with $n$ depends on the width of the wave packets.
Ten significant figure accuracy was achieved 
in the test model of ref (\cite{Kopp:2011vv}).

The third approximation that we used was to uniformly approximate
$e^{2ine^{-\beta H}}$ by a polynomial in $e^{-\beta H}$.  This was
possible because the spectrum of $e^{-\beta H}$ is bounded.  
Because of the uniform convergence the error is identical to 
the error in approximating 
\beq
\vert e^{2inx} - P(x) \vert < \epsilon  \qquad x \in [0,1]
\eeq
by a polynomial.  Ten significant figure accuracy was again realized using 
Chebyshev polynomials with Gauss-Chebyshev quadratures.  This method
requires that the approximations be performed in the specified order.

In more realistic models an additional approximation is needed,
which is the solution of point spectrum mass eigenstates of the
subsystem Green functions.  These appear in the multi-channel
generalization of the mapping $\Phi$ and are needed to get the strong
convergence needed to satisfy the Cook condition (\ref{i.20}).
They can also appear in the two-point function if it has a 
non-trivial Lehmann weight. 

In general, given an explicit Hilbert space representation and knowing
that the scattering theory exists, there are many other techniques
that could be used to calculate scattering observables without
analytic continuation.  The method discussed above provides one 
method that has been tested, but it may not be the most efficient 
method available.

\section{QCD}

Ultimately one would like to use Euclidean methods to compute GeV scale
scattering observables in QCD.  Lattice, path integral, and Schwinger-Dyson 
formulations of QCD all yield Euclidean Green functions.

In QCD, because of confinement, the Euclidean Green functions of the
theory are not expected to be reflection positive.  However reflection
positivity should hold for color singlet initial and final states.  In
addition the scattering asymptotic states should also be reflection
positive.  The Euclidean methods discussed in this paper are still
be applicable if these two conditions hold.

\section{Summary} 

In this paper the existence of non-trivial reflection positive
Euclidean Green functions was demonstrated by exhibiting an explicit
class of reflection positive connected four-point Euclidean Green
functions.  The structure of this class of Green functions was
motivated by a theorem of Widder that exhibited the structure of a
general one-dimensional reflection-positive two-point function.  The
general structure of reflection positive four-point functions is still
an open problem.  More importantly, the structure of Euclidean Green
functions for realistic models remains an open problem.

In this paper time-dependent scattering methods were used to
demonstrate the existence of scattering wave operators for models
based on the reflection positive Euclidean Green functions of the form
(\ref{f.8}).  The basic observation is that the Cook condition that is
normally used as a sufficient condition for the existence of
non-relativistic wave operators can be applied in this formulation of
Euclidean relativistic quantum mechanics.  The $t^{-3/2}$ asymptotic
behavior of the integrand in (\ref{i.20}) that ensures the existence of
the wave operator for short-ranged potentials in the non-relativistic
case is realized in the relativistic case for sufficiently 
well-behaved connected Euclidean Green function functions.

The results of this paper imply that the Euclidean methods tested in
\cite{Kopp:2011vv}, when applied to models defined by the class of
reflection positive Green function in section three, should converge
to transition matrix elements for a range of energies up to the few
GeV scale.

This work was supported by the U.~S.
Department of Energy, Office of Nuclear Physics, under contract
No. DE-FG02-86ER40286.

\section{appendix - spin}

Two point Euclidean Green functions that lead to any positive-mass
irreducible representation space of the Poincar\'e group are
constructed in this appendix.

A basis for vectors in a positive-mass irreducible representation
space of the Poincar\'e group consists of simultaneous eigenstates of
the mass, spin, linear momentum, and $z$-component of some
kind of spin (canonical, Jacob-Wick helicity, light-front, $\cdots$).  
These states have the following transformation property
\[
U(\Lambda ,a) \vert (m,j) \mathbf{p}, \mu \rangle =
\]
\beq
\sum_{\nu} e^{-i \Lambda p \cdot a}  \vert \pmb{\Lambda} p , \nu
\rangle
D^j_{\nu \mu} [ B^{-1}(\Lambda p/m) \Lambda B(p/m) ]
\sqrt{{\omega_m (\Lambda p) \over \omega_m(p)}}
\label{k.1}
\eeq
where $B^{-1}(\Lambda p/m)\Lambda B(p/m)$ is a Wigner rotation.
The choice of Lorentz boost, $B(p/m)^{\mu}{}_{\nu}$, in the 
Wigner rotation 
determines the type of spin \cite{Keister:1991sb}.  For any kind of spin 
the Wigner $D$ functions, which are also finite dimensional 
representations of $SL(2,\mathbb{C})$,  can be factored into products.
Multiplication of both sides of (\ref{k.1}) by 
\beq
D^j_{\mu \nu} [ B^{-1}(p/m) ]
\eeq
leads to 
\[
\sum_{\mu}  U(\Lambda ,a) \vert (m,j) \mathbf{p}, \mu \rangle
D^j_{\mu \sigma} [ B^{-1}(p/m)] 
\sqrt{\omega_m (p)}=
\]
\beq
\sum_{\nu \sigma'} e^{-i \Lambda p \cdot a}  \vert \pmb{\Lambda} p , \nu
\rangle
D^j_{\nu \sigma'} [ B^{-1}(\Lambda p/m)]  
\sqrt{\omega_m (\Lambda p)}
D^j_{\sigma' \sigma} [\Lambda  ].
\label{k.2}
\eeq
The vectors
\beq
\vert p ,j , \sigma \rangle := 
\sum_{\mu} \vert (m,j) \mathbf{p}, \mu \rangle
D^j_{\mu \sigma} [ B^{-1}(p/m)] 
\sqrt{\omega_m (p)}
\label{k.3}
\eeq
transform in a Lorentz covariant manner
\beq
U(\Lambda, 0) \vert p ,j , \sigma \rangle =
\sum_{\sigma'}e^{-i \Lambda p \cdot a}
\vert \Lambda p ,j , \sigma' \rangle
D^j_{\sigma' \sigma} [\Lambda  ].
\label{k.4}
\eeq
The transformation $U(\Lambda, 0)$
is unitary with respect to the inner product
\beq
\psi(p,j,\sigma) = 
\langle p ,j , \sigma \vert \psi \rangle ,
\label{k.5}
\eeq
\beq
\langle \psi \vert \phi \rangle = 
\sum_{\mu \sigma} \int
\psi^*(p,j,\mu)  D^j_{\mu \sigma} [ B(p/m) B^{\dagger}(p/m)]  {md\mathbf{p}
\over  \omega_m (p) } 
\phi (p,j,\sigma)  
\label{k.6}
\eeq
where $p_0=\omega_m(\mathbf{p})$ is the energy.
The kernel simply removes the momentum-dependent $SL(2,C)$
Wigner functions from the covariant representation.  Because 
the $SL(2,\mathbb{C})$ matrices cancel in computing matrix elements
- the result is the same independent of whether the right or 
left-handed representations 
of $SL(2,\mathbb{C})$ are used.

Note that in $SL(2,\mathbb{C})$ a general boost has a polar decomposition
\beq
B(p) = P(p) R(p)  
\label{k.7}
\eeq
where $P(p)$ is the positive Hermitian operator, 
\beq
P(p) = e^{ \pmb{\rho}\cdot \pmb{\sigma} /2},
\label{k.8}
\eeq 
$\pmb{\rho}$ is the rapidity vector and $R(p)$ is an $SU(2)$
matrix  (generalized Melosh rotation) that determines the type of spin. 
It follows that 
\beq
B(p/m) B^{\dagger}(p/m) = P(p) R(p) R^{\dagger}(p) P(p )=
P^2 (p) = e^{\pmb{\rho} \cdot \pmb{\sigma}} = \sigma \cdot p .
\label{k.9}
\eeq
In this expression the Melosh rotations cancel, so the result is
independent of the choice of spins. 
Thus this scalar product can be expressed as
\beq
\sum_{\alpha \beta}\int
\psi^*(p,j, \alpha)  D^j_{\alpha \beta} [ p \cdot \sigma ] m {d\mathbf{p}
\over 
\omega_m (p) } 
\phi (p,j,\beta) = 
\label{k.10}
\eeq
\beq 
\sum_{\alpha \beta} \int
\psi^*(p,j,\alpha)  D^j_{\alpha \beta } [ p \cdot \sigma ] 2m d^4{p} 
\delta (p^2 + m^2)  
\phi (p,j,\beta). 
\label{k.11}
\eeq
This is essentially identical to the form found in \cite{Wightman:1980}
(see eq. 1.57).
The important observation is that $\sigma \cdot p$ is a positive 
Hermetian matrix for timelike $p$.  The same holds for 
$D^j_{\mu \sigma} [ p \cdot \sigma^t ]$,
$D^j_{\mu \sigma} [ p \cdot \sigma^* ]$, and
$D^j_{\mu \sigma} [ p \cdot \sigma^{-1}]$. 

The following Green function is a Euclidean 
covariant rather than Euclidean invariant distribution
\beq
\int {2m D^j_{\alpha \beta } [ p_e \cdot \sigma_e ] \over
p_e^2 +m^2 } 
d^4 p_e e^{i p_e \cdot (\x-\y)}
\label{k.12}
\eeq
that leads exactly to the representation (\ref{k.6}) of 
a mass $m$ spin $j$ irreducible representation.

These considerations show that the following Euclidean scalar product is 
reflection positive  
\[
\sum_{\alpha \beta}
\int g^*_\alpha (\theta \x) 
{2m D^j_{\alpha \beta } [ p_e \cdot \sigma_e ] \over
p_e^2 +m^2 } 
d^4 p_e e^{i p_e \cdot (\x-\y)} g_\beta (\y)d^4x d^4y d^4p_e =
\]
\beq
\sum_{\alpha \beta}
\psi^*(p,j,\alpha)  D^j_{\alpha \beta } [ p \cdot \sigma ] 2m d^4{p} 
\delta (p^2 + m^2)  
\psi (p,j,\beta)
\label{k.13}
\eeq
with 
\beq
\psi (p,j,\beta) = 
\int g_\beta (\x^0,\mathbf{\x} )e^{-\omega_m (\mathbf{p}) \x^0}
e^{i \mathbf{p}\cdot \mathbf{\x}}.
\label{k.14}
\eeq
This shows how to construct reflection-positive two-point
Euclidean Green 
functions for any irreducible representation of the Poincar\'e group.
While I did not choose to double the representation, doubled
representations can be realized by replacing
$
D^j_{\alpha \beta } [ p \cdot \sigma ] 
$
by a direct sum of a right and left handed representation ,
$
D^j_{\alpha' \beta' } [ p \cdot \sigma_2  \sigma^* \sigma_2  ] 
$, 
which is also positive for positive energy timelike $p$.


\end{document}